\documentclass[
    ,final            
  ]
  {aipproc}

\layoutstyle{6x9}

\begin{document}
\newcommand{\newc}{\newcommand}
\newc{\ra}{\rightarrow}
\newc{\lra}{\leftrightarrow}
\newc{\beq}{\begin{equation}}
\newc{\eeq}{\end{equation}}
\newc{\barr}{\begin{eqnarray}}
\newc{\earr}{\end{eqnarray}}
\newcommand{\Od}{{\cal O}}
\newcommand{\lsim}   {\mathrel{\mathop{\kern 0pt \rlap
  {\raise.2ex\hbox{$<$}}}
  \lower.9ex\hbox{\kern-.190em $\sim$}}}
\newcommand{\gsim}   {\mathrel{\mathop{\kern 0pt \rlap
  {\raise.2ex\hbox{$>$}}}
  \lower.9ex\hbox{\kern-.190em $\sim$}}}

\title{SOLAR NEUTRINOS AS   BACKGROUND IN DIRECT DARK MATTER SEARCHES.}

\classification{13.15.+g, 14.60Lm, 14.60Bq, 23.40.-s, 95.55.Vj, 12.15.-y.}
\keywords      {WIMPs, LSP, Dark Matter, Direct detection, Boron neutrinos, Quenching factor,
Differential rates, Total rates, Neutral current, Neutron coherence}

\author{J. D. Vergados}{
  address={ Physics Department, University of Ioannina, Ioannina, GR 451 10, Greece and\\
  Theory Division, CERN 1211, Geneva 23, CH.}
}
\author{H. Ejiri}{address={ RCNP, Osaka University, Osaka, 567-0047, Japan and\\
 National Institute of Radiological Sciences, Chiba, 263-8555, Japan.}}
\author{I. Giomataris}{
  address={IRFU, Centre d' etudes de Saclay, 91191 Gif sur Yvette CEDEX,  France.}
}
%

\begin{abstract}
The coherent contribution of all neutrons in neutrino nucleus scattering due to the neutral current is examined considering the boron solar neutrinos. These neutrinos could potentially become a source of background in the future dark matter searches aiming at nucleon cross sections in the region well below the $10^{-10}$pb, i.e a few events per ton per year.
\end{abstract}

\maketitle


\section{Introduction.}
The universe is observed to contain large amounts of dark matter
\cite{WMAP06,SPERGEL}, and its contribution to the total energy
density is estimated to be $\sim 25\%$. 
Even though there exists firm indirect
evidence from the halos of dark matter in galaxies and clusters of
galaxies it is essential to detect matter directly.

The possibility of direct detection, however, depends on the nature of
the dark matter constituents, i.e the WIMPs (weakly interacting massive particles).
 Supersymmetry naturally provides
candidates for these constituents \cite{ALLSUSY}.
In the most favored
scenario of supersymmetry, the lightest supersymmetric particle
(LSP) can be described as a Majorana fermion, a linear combination
of the neutral components of the gauginos and higgsinos. Other
possibilities also exist, see, e.g. some models in universal
theories with extra dimensions \cite{SERVANT},\cite{OikVerMou}.
Since the  WIMPs are expected to be very massive ($m_{{WIMP}} \geq
30$ GeV) and extremely non-relativistic with average kinetic
energy $ \langle T \rangle \simeq$ 50 keV $\left(m_{{WIMP}}/ 100\,
{\rm GeV} \right)$, a WIMP interaction with a nucleus in an
underground detector is not likely to produce excitation.  As a
result, WIMPs can be directly detected mainly via the recoil of a
nucleus ($A,Z$) in elastic scattering. The event rate for such a
process can be computed  by a standard procedure (see recent work
\cite{JDVEJ08} and references therein). 
 Since the particle
physics parameters will most likely result in very small cross
sections, the future dark matter experiments, like
the XENON-ZEPLIN \cite{CDMALL}, aim at detecting 10 events per ton per year. At
this  level one may encounter very bothersome backgrounds. One
such background may come from  the high energy boron solar
neutrinos \cite{JDVEJ08}, \cite{MonFis}
 (the other neutrinos are characterized either by too
small energy or much lower fluxes).

 During the last years various detectors aiming at detecting recoiling nuclei have been
 developed
in connection with  dark matter
searches \cite{CDMALL} with thresholds in the few keV region. Furthermore it has become feasible to 
detect neutrinos by measuring
 the recoiling nucleus  in gaseous detectors with much lower threshold energies \cite{VERGIOM06}, explpoiting 
the neutral current interaction \cite{DKLEIN}.
This interaction, through its vector component, can lead
 to coherence, i.e. an additive contribution of all neutrons in the nucleus.

 In this paper
 we will calculate the recoil spectrum 
  for boron solar neutrinos and compare
 it with that
associated with WIMPs, both for a light and a heavy target.

  \section{A brief discussion of the rates for direct WIMP detection}
   Before proceeding with the evaluation of the event rate for nuclear recoils originating from
   the neutrino nuclear scattering we will briefly summarize the results recently obtained \cite{JDVEJ08}.
The  event rate for the coherent WIMP-nucleus elastic scattering, which is given by:
\begin{eqnarray}
R&=& \frac{\rho (0)}{m_{\chi^0}} \frac{m}{m_p}~
              \sqrt{\langle \upsilon^2 \rangle }\sigma_{p,\chi^0}^{S}f_{coh}(A,\mu_r(A))
\nonumber\\
& &f_{coh}(A,\mu_r(A))=
              \frac{100\mbox{GeV}}{m_{\chi^0}}\left[ \frac{\mu_r(A)}{\mu_r(p)}
 \right]^2 A~t_{coh}\left(1+h_{coh}cos\alpha \right)
\label{fullrate}
\end{eqnarray}
where $ \sigma_{p,\chi^0}^{S} $ is the coherent WIMP-nucleon cross section, $\mu_r(A) $ the WIMP-nucleus
reduced mass 
  The
parameter $t_{coh}$ takes into account the folding of the nuclear form factor with the WIMP
 velocity distribution, $h_{coh}$ deals with the modulation \cite{JDV03} due to the motion of the Earth and $\alpha$ is
 the phase of the Earth (zero around June 2nd).

\begin{figure}[!ht]
\rotatebox{90}{\hspace{-0.0cm} {$\frac{dR_{coh}}{dQ}$  $\rightarrow keV^{-1} \frac{\sigma_{p,\chi^0}^{S}}{10^{-6}pb}$}}
\includegraphics[height=0.14\textheight]{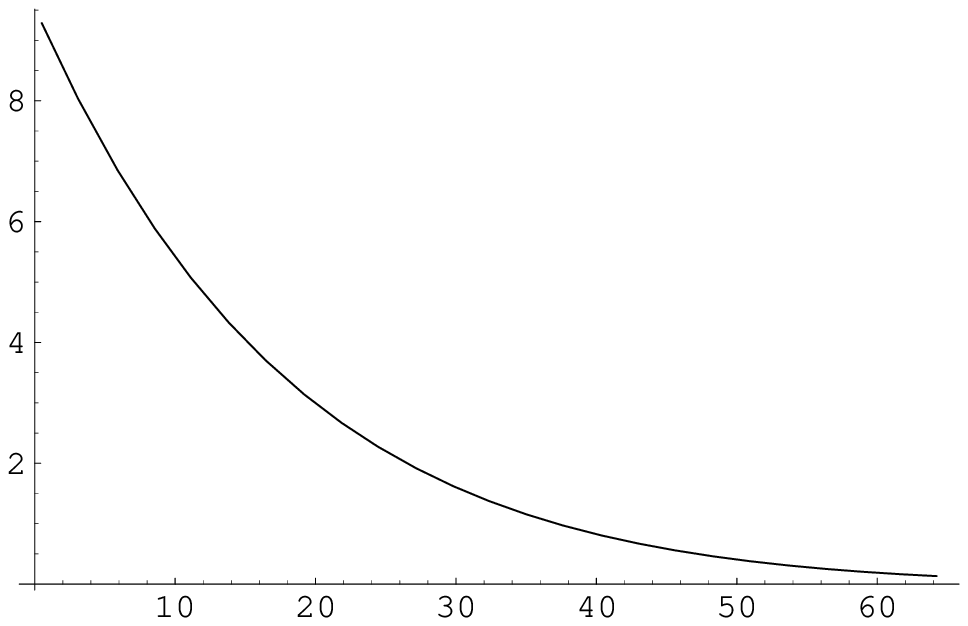}\\
{\hspace{0.0cm} $Q \rightarrow$  keV}
\includegraphics[height=0.14\textheight]{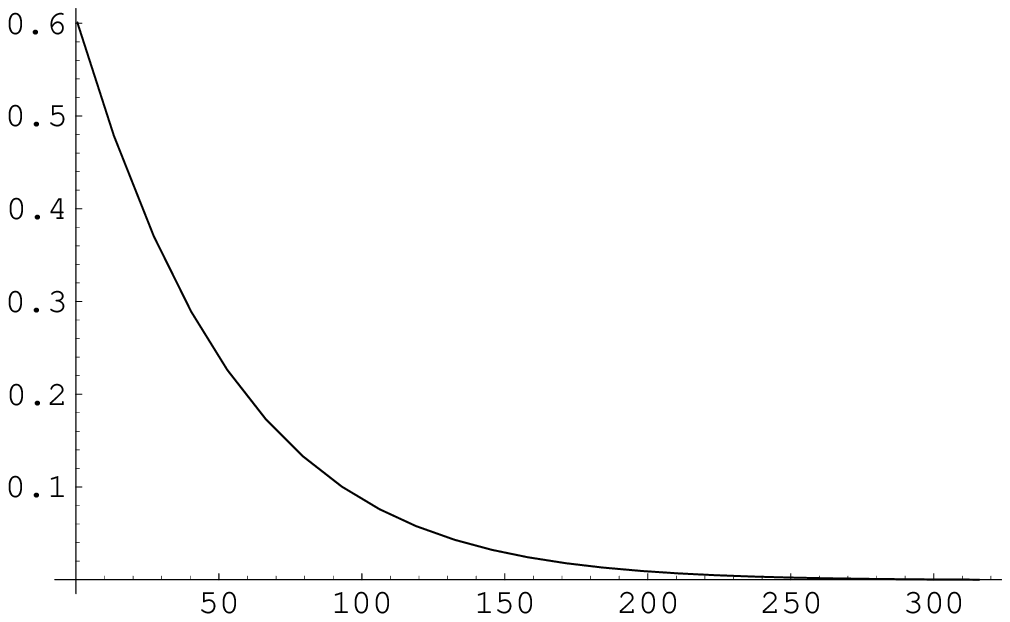}\\
{\hspace{0.0cm} $Q \rightarrow$  keV}
 \caption{We  show the differential event rate  $ {dR_{coh}}/{dQ}$ for the coherent process , as a function of the energy transfer,
   for  a WIMP mass, $m_{\chi}$, of 100 GeV
in the case of $^{131}$Xe on the left and $^{32}$S on the right.}
\label{fig:drdQ127}
  \end{figure}

\begin{figure}[!ht]
\rotatebox{90}{\hspace{-0.0cm} {Event rate per kg-yr
$\rightarrow \frac{\sigma_{p,\chi^0}^{S}}{10^{-6}pb}$}}
\includegraphics[height=0.13\textheight]{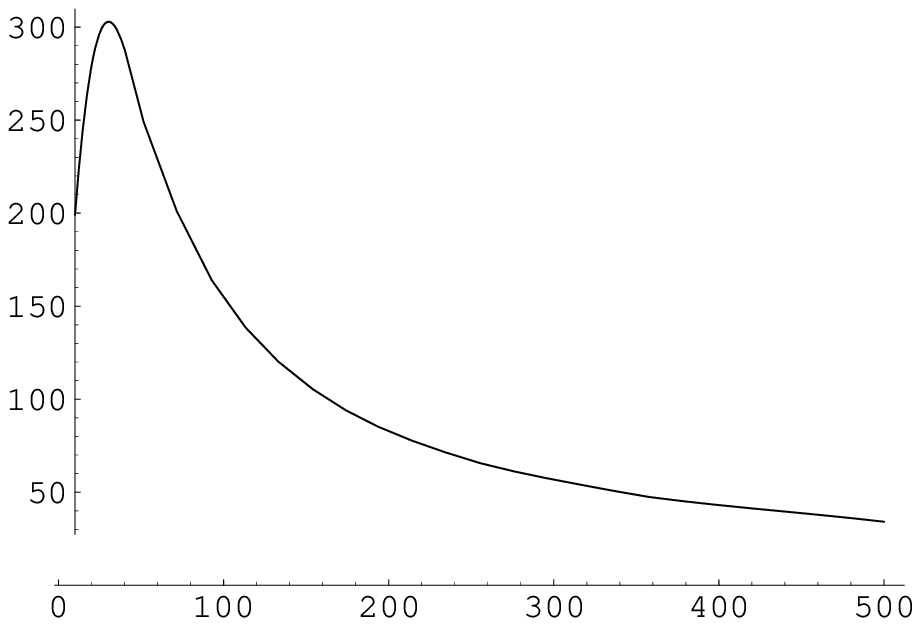}\\
{\hspace{0.0cm} $m_{\chi} \rightarrow$ GeV }
\includegraphics[height=0.13\textheight]{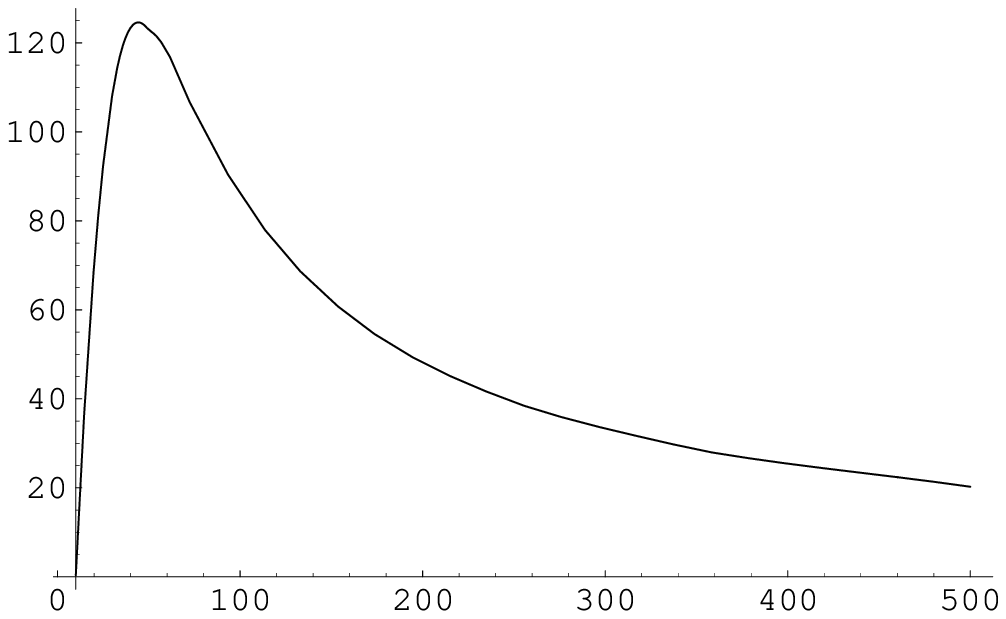}\\
{\hspace{0.0cm} $m_{\chi} \rightarrow $ GeV }
 \caption{We  show the total event rate   for the coherent process  as a function of the  WIMP mass in the case of $^{131}$Xe for zero threshold on the left ans 10 keV on the right.}
 \label{fig:rate131}
  \end{figure}

 The number of events in time $t$ due to the scalar interaction, which leads to coherence  \cite{JDVEJ08}, is:
\beq
 R\simeq  1.60~10^{-3}\times\frac{t}{1 \mbox{y}} \frac{\rho(0)}{ \mbox {0.3GeVcm}^{-3}}
\frac{m}{\mbox{1Kg}}\frac{ \sqrt{\langle
\upsilon^2 \rangle }}{280 {\mbox {km s}}^{-1}}\frac{\sigma_{p,\chi^0}^{S}}{10^{-6} \mbox{ pb}} f_{coh}(A, \mu_r(A))
\label{eventrate}
\eeq
  Assuming a constant nucleon cross section we get the differential rate
 indicated in Fig. \ref{fig:drdQ127}.
   The total (time averaged) coherent event rate is shown in
Fig. \ref{fig:rate131}.

\section{ Elastic Neutrino nucleon Scattering}
The cross section for elastic neutrino nucleon scattering has extensively been studied.
It has been shown that at low energies it can be simplified and  be cast in the form:
\cite{BEACFARVOG},\cite{VogEng}:
 \beq
 \left(\frac{d\sigma}{dT_N}\right)_{weak}=\frac{G^2_F m_N}{2 \pi}
 [(g_V+g_A)^2 + (g_V-g_A)^2 [1-\frac{T_N}{E_{\nu}}]^2
+ (g_A^2-g_V^2)\frac{m_NT_N}{E^2_{\nu}}]
 \label{elasw}
  \eeq
  where $T_n$ is the energy of the nucleon, $m_N$ the nucleon mass and $g_V$, $g_A$ are the weak coupling constants:
  \beq
 g_V=-2\sin^2\theta_W+1/2\approx 0.04~,~g_A=\frac{1.27}{2} ~~,~~(\nu,p)~~;~~g_V=-1/2~,~g_A=-\frac{1.27}{2}~~,~~(\nu,n)
\label{gcoup1}
\eeq
 
 The elastic neutrino nucleus scattering is given by:
\beq
\left(\frac{d\sigma}{dT_A}\right)_{weak}=\frac{G^2_F Am_N}{2 \pi}~(N^2/4) F^2(q^2)
  \left ( 1+(1-\frac{T_A}{E_{\nu}})^2
-\frac{Am_NT_A}{E^2_{\nu}} \right)
 \label{elaswAV1}
\eeq
with 
 $T_A$ is the energy of the recoiling nucleus and
 $F(q^2)= F(T_A^2+2 A m_N T_A)$

\begin{figure}[!ht]
 \rotatebox{90}{\hspace{1.0cm} {$f_{\nu}(E_{\nu})\rightarrow $}}
\includegraphics[scale=0.8]{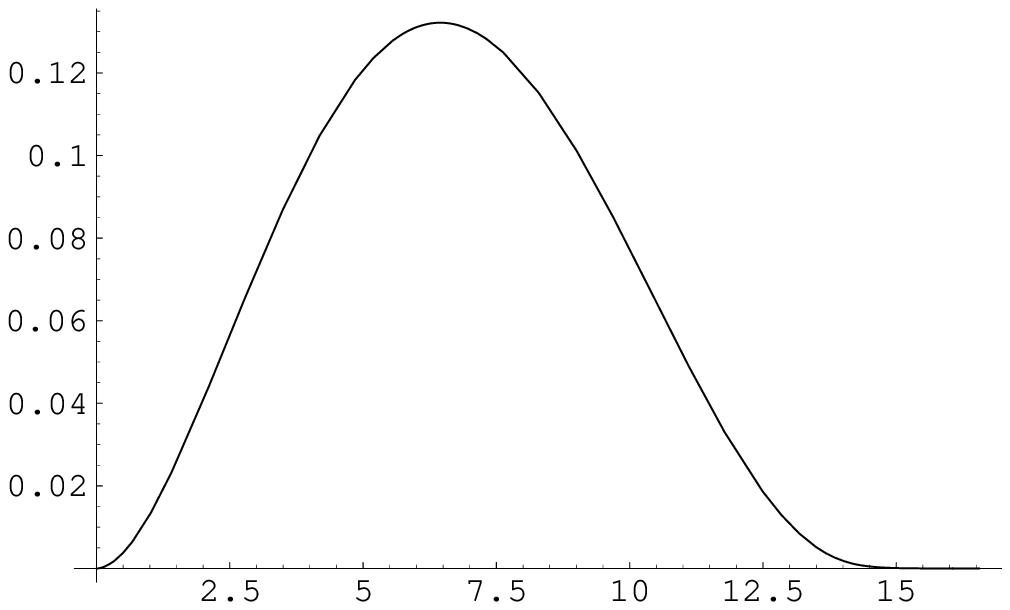}\\
\hspace{0.0cm}$E_{\nu} \rightarrow$ MeV
 \caption{The boron solar neutrino spectrum. The total neutrino flux is $\Phi_{\nu}=5.15 \times 10^6$cm$^2$ s $^{-1} $.}
 \label{fig:nuspec}
  \end{figure}

\begin{figure}[!ht]
 \rotatebox{90}{\hspace{0.0cm} {$\frac{d\sigma}{dT_A} \rightarrow 10^{42}$ cm$^2$ keV$^{-1}$ }}
\includegraphics[scale=0.5]{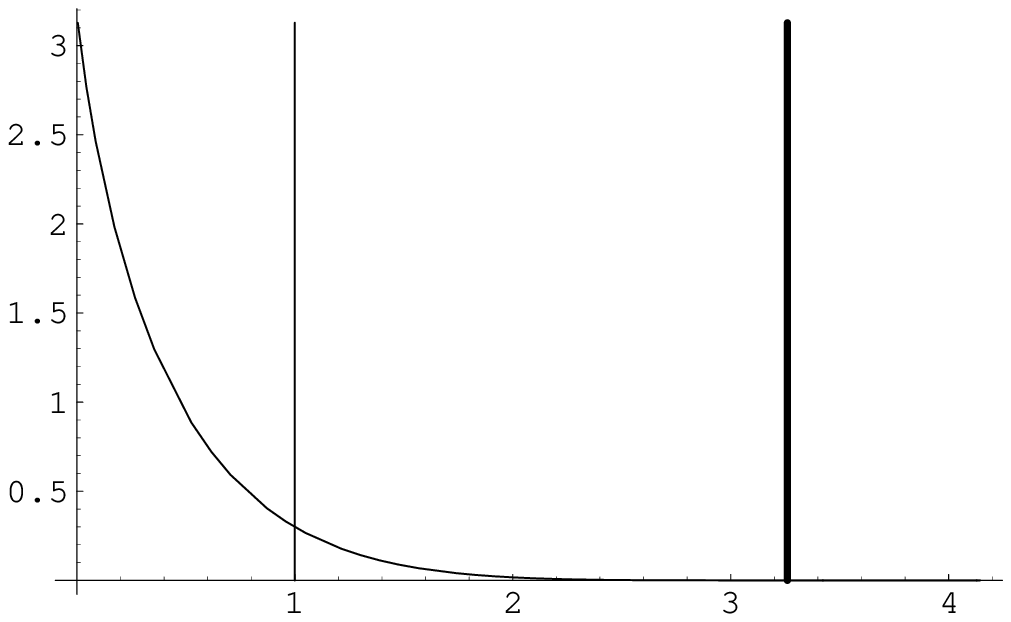}\\
\hspace{0.0cm}$T_A \rightarrow keV$
\includegraphics[scale=0.5]{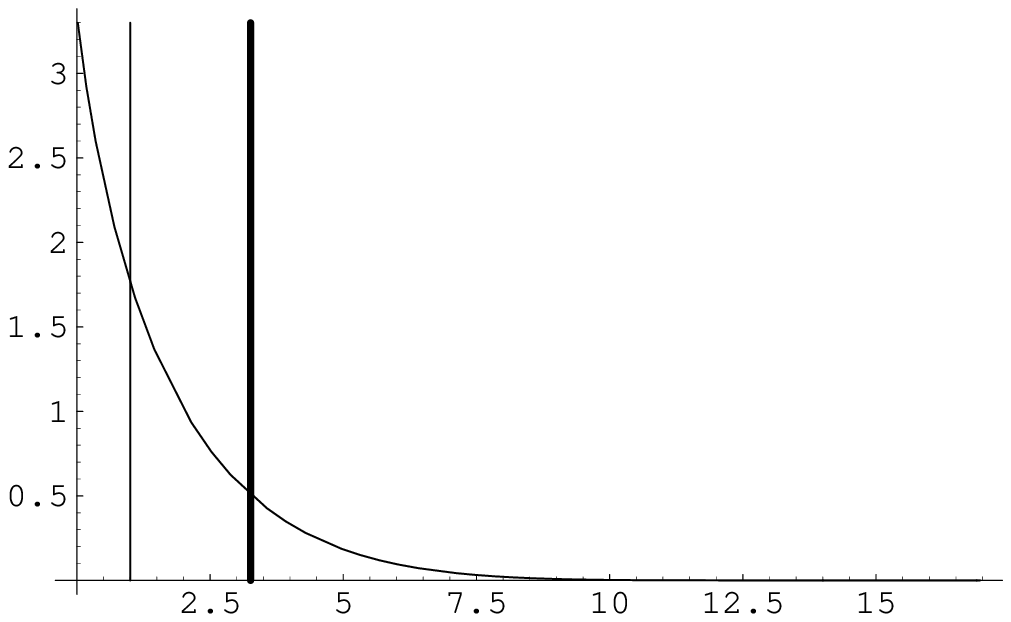}\\
\hspace{0.0cm}$T_A \rightarrow keV$
 \caption{The neutrino induced differential cross section in units $10^{-42} $cm$^2$ keV$^{-1}$
as a function of the recoil energy in keV
in the case of the target $^{131}$Xe on the left and $^{32}$S on the right. Note that essentially all the contribution comes from the
 recoil energy region $0\leq T_A\leq 1$ keV. The available kinematic region for a threshold energy of 1 keV is that 
which is above the thin line (thick line) for no quenching (quenching) respectively.
The effect of the form factor is invisible in the figure.}
 \label{fig:dsigma131}
  \end{figure}

  To proceed further we must convolute the cross section with the neutrino spectrum.
For the relwvant boron solar  neutrinos 
the normalized spectrum is shown in
  Fig. \ref{fig:nuspec}. The expected total neutrino flux is $\Phi_{\nu}=5.15 \times 10^6$cm$^2$ s $^{-1} $.
The obtained differential cross section is shown in Fig.  \ref{fig:dsigma131}.
Note that the cross section has been almost completely depleted beyond energies 1 and 5 keV for a
intermediate ($^{131}$Xe ) and light ($^{32}$S) targets respectively.
  Integrating the differential cross section down to zero threshold  we find the  event rates given in table
\ref{table.nurates}. The event rates are almost two orders of magnitude smaller than the rates
for WIMP detection obtained with a nucleon cross section of $10^{-9}$pb. Thus such neutrinos cannot be a serious background for WIMP searches in the region $10^{-9}-10^{-10}$pb. In any event, as we will see below, the neutrino induced recoils are less of a background problem
in the realistic case of non zero energy threshold.
\begin{table}[t]
\caption{ Comparison of the event rates for boron solar  neutrino and  WIMP
detection. In evaluating the latter we assumed a nucleon cross section independent of the mass. The
kinematics were obtained assuming two WIMP masses, namely 100 and 300 GeV. NoFF means that the nuclear form factor
was neglected.
}
\label{table.nurates}
\begin{tabular}{|c|c|c|c|c|}
\hline
   &  &  & &\\
target &$R_{\chi}$(kg-y)$\times \frac{\sigma_N}{10^{-9}pb}$&$R_{\chi}$(kg-y)$\times  \frac{\sigma_N}{10^{-9}pb}$&$R_{\nu}$(kg-y)&$R_{\nu}$(kg-y);NoFF\\
\hline
&$m_{\chi}=100$GeV&$m_{\chi}=300$GeV& \\
\hline
$^{131}$Xe&0.167&0.060&0.934$\times 10^{-3}$&0.952$\times 10^{-3}$\\
$^{32}$S&0.033&0.014&0.167$\times 10^{-3}$&0.168$\times 10^{-3}$\\
\hline
\end{tabular}
\end{table}

\begin{figure}[!ht]
 \rotatebox{90}{\hspace{1.0cm} {$\frac{\sigma(E_{thr})}{\sigma(E_{thr}=0)} \rightarrow $}}
\includegraphics[scale=0.5]{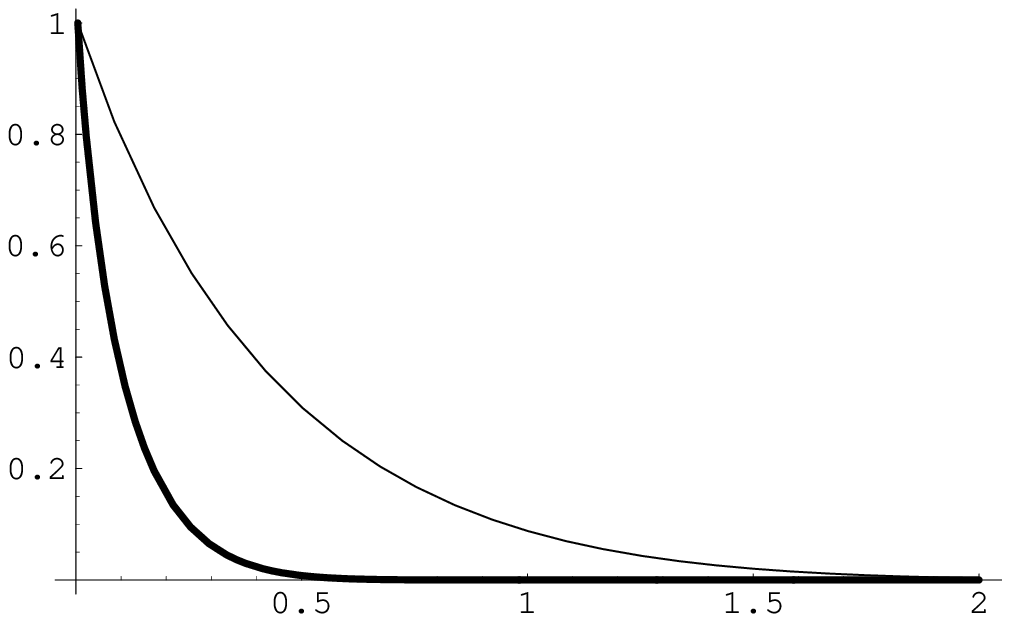}\\
\hspace{0.0cm}$ E_{thr}\rightarrow $ keV
\includegraphics[scale=0.5]{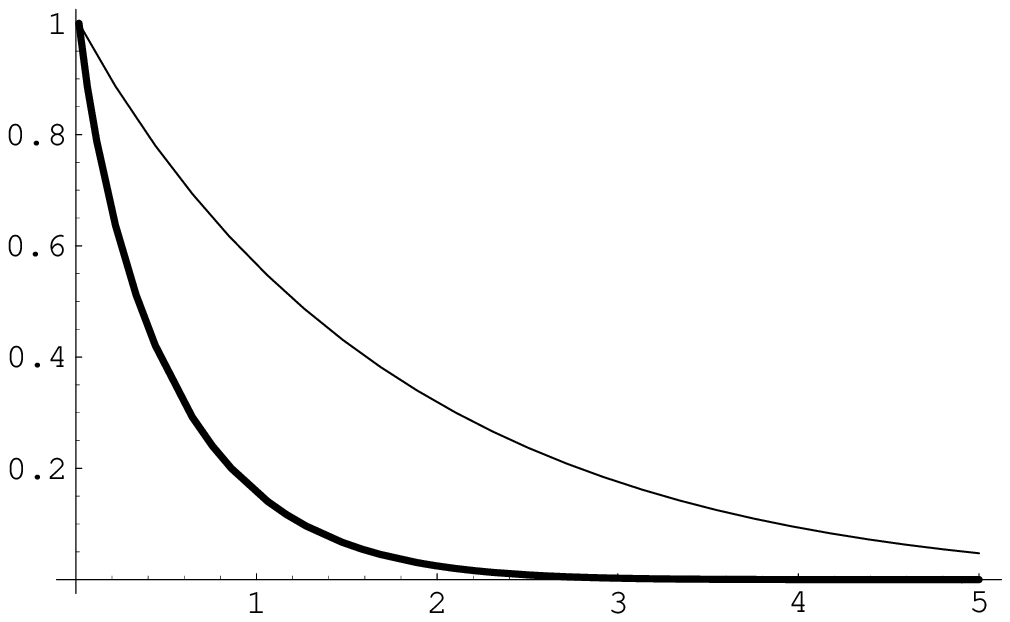}\\
\hspace{0.0cm}$  E_{thr}\rightarrow$ keV
 \caption{The ratio of the total cross section with threshold divided by that with zero threshold   respectively. The thick (thin) line correspond to quenching (no quenching). Note  that
 the observed events are an order of magnitude down,  if the energy threshold is 1 keV. }
 \label{fig:thr131}
  \end{figure}

We should mention that the obtained rates are independent of the neutrino oscillation parameters, since the neutral
current events are not affected by such oscillations.
The above results shown in table  \ref{table.nurates}), refer to an ideal detector with zero energy threshold. 
For a real detector, however,
 the nuclear recoil events are quenched, especially at low energies.
The quenching
factor for a given detector  is the ratio of the signal height for a recoil track divided by that
 of an electron signal height with the same energy.  The actual quenching
  factors must be determined experimentally for each target. In the present work we considered \cite{JDVEJ08} the
phenomenological factor :
\beq
Q_{fac}(T_A)=r_1\left[ \frac{T_A}{1keV}\right]^{r_2},~~r_1\simeq 0.256~~,~~r_2\simeq 0.153
\label{quench1}
\eeq
   Due to the relatively low recoil energies the
effect of threshold is crucial (see Fig  \ref{fig:thr131}). One clearly sees that
 the observed events are an order of magnitude down  if the energy threshold is 1 keV (2 keV) for Xe(S)
respectively.  Thus with quenching most
  signals get below the threshold energy of $\approx$ 1 keV.
On the other hand the WIMP
event rates are almost unaffected, unless the threshold energy becomes larger than 5 keV.

  \section{Concluding remarks}
  In the present study we considered the elastic scattering of WIMP-nucleus interaction and the
corresponding elastic scattering of boron solar neutrinos.
Our results can be summarized as follows:
\begin{enumerate}
\item The differential cross section for solar neutrinos decreases sharply as the nuclear recoil energy increases. It almost vanishes beyond 1 keV (5 keV) for intermediate (light target), like $^{131}Xe$ ($^{32}S$). On the other
the corresponding event rates for WIMPs of mass $\approx$ 100 GeV extend further than 30 (150) keV $^{131}Xe$ ($^{32}S$) respectively.
\item The event rates for boron solar neutrinos at zero threshold energy and no quenching are 2-3
orders of magnitude smaller than those for WIMPs with a nucleon cross section $10^{-9}$pb. Thus
solar neutrinos are not a serious background down to $10^{-10}$pb.
\item Since the nuclear recoil energy in solar neutrino scattering is smaller than that
associated with heavy WIMPs, one  can further
substantially decrease its contribution by restricting the observation
 above a few keV without seriously affecting the corresponding WIMP rates.
Thus neutrinos do not appear to be a serious background even at the level of $10^{-11}$pb.
\item One may reduce this background still further by exploiting the quenching factors.
\item One may diminish this bakground further by exploiting the annual modulation of the signal (see e.g. \cite{JDV03} and references there in)  or even better by performing
 directional experiments \cite{DRIFT},\cite{VF07},
   in which the direction of recoil is also measured, one will be able to select WIMP signals
and discriminate against neutrino scattering.
\end{enumerate}
\begin{theacknowledgments}
 One of the authors (JDV)  is indebted to MRTN-CT-2004-503369 for support and financing his participation in DSU2008.
 Special thanks to  BUE and professor S. Khalil for their hospitality as well as to the CERN Theory Division for support and hospitality.
\end{theacknowledgments}

\end{document}